\documentclass[number]{elsarticle}

\usepackage[T1]{fontenc}
\usepackage[utf8]{inputenc}
\usepackage{subfig}
\usepackage{amsmath}
\usepackage{amssymb}
\usepackage{array}
\usepackage{listings}
\usepackage{algorithm}
\usepackage{algpseudocode}
\usepackage{float}
\usepackage{floatpag}
\usepackage{siunitx}
\usepackage{url}
\usepackage{verbatim}
\usepackage{setspace}
\usepackage{hyperref}

\newcolumntype{L}[1]{>{\raggedright\let\newline\\\arraybackslash\hspace{0pt}}m{#1}}
\newcolumntype{R}[1]{>{\raggedleft\let\newline\\\arraybackslash\hspace{0pt}}m{#1}}

\journal{Journal of Parallel and Distributed Computing}

\begin{document}

\begin{frontmatter}

\title{Locality Optimized Unstructured Mesh Algorithms\\ on GPUs}

  \author[1,2]{A.A. Sulyok\corref{corres}}
  \ead{sulyok.andras.attila@itk.ppke.hu}
  \author[1,2]{G.D. Balogh}
  \author[1]{I.Z. Reguly}
  \author[3]{G.R. Mudalige}
  \address[1]{
    Faculty of Information Technology and Bionics,
    Pázmány Péter Catholic University,
    Budapest, Hungary
  }
  \address[2]{
    3in-PPCU Research Group,
    Pázmány Péter Catholic University,
    Esztergom, Hungary
  }
   \address[3]{
    Department of Computer Science,
    University of Warwick,
    Coventry, United Kingdom
  }
  \cortext[corres]{Corresponding author}

\begin{abstract}

\noindent Unstructured-mesh based numerical algorithms such as finite volume and 
finite element algorithms form an important class of applications for many 
scientific and engineering domains. The key difficulty in achieving higher 
performance from these applications is the indirect accesses that lead to 
data-races when parallelized. Current methods for handling such data-races 
lead to reduced parallelism and suboptimal performance. Particularly on modern 
many-core architectures, such as GPUs, that has increasing core/thread 
counts, reducing data movement and exploiting memory locality is vital for 
gaining good performance.  

In this work we present novel locality-exploiting optimizations for the 
efficient execution of unstructured-mesh algorithms on GPUs. Building on a 
two-layered coloring strategy for handling data races, we introduce novel 
reordering and partitioning techniques to further improve efficient execution. 
The new optimizations are then applied to several well established 
unstructured-mesh applications, investigating their performance on NVIDIA's 
latest P100 and V100 GPUs. We demonstrate significant speedups 
($1.1\text{--}1.75\times$) compared to the state-of-the-art. A range of 
performance metrics are benchmarked including runtime, memory transactions, 
achieved bandwidth performance, GPU occupancy and data reuse factors and 
are used to understand and explain the key factors impacting performance. The 
optimized algorithms are implemented as an open-source software library and 
we illustrate its use for improving performance of existing or new 
unstructured-mesh applications.
\end{abstract}

\begin{keyword}
finite volume \sep finite element \sep race condition \sep GPU
\end{keyword}

\end{frontmatter}

\section{Introduction}\label{introduction}

\noindent Unstructured mesh solvers, particularly applied to the solution of 
finite difference, finite volume or finite element algorithms, form the basis 
of numerical simulation applications in a vast area of important scientific 
domains, from modeling the flow of blood in the body, the flow past an aircraft, 
to ocean circulation and the simulation of Tsunamis. Significant computational 
resources are required for the execution of numerical algorithms on these highly 
detailed (usually three-dimensional) meshes. The solution involves repeatedly 
iterating over millions of elements (such as mesh edges, nodes, etc.) to reach 
the desired accuracy or resolution. The key distinguishing feature of these 
applications is that operations over mesh elements make use of explicit 
connectivity information between elements to access data defined on neighboring 
elements. This is in contrast to the use of stencils in structured-mesh 
applications where the regular geometry of the mesh implicitly provides the 
connectivity information. As such, iterations over unstructured-meshes lead to 
highly irregular patterns of data accesses over the mesh, characterized by 
indirect array accesses. For example, computations over the mesh involve 
iterating over elements of a set (e.g. faces), performing the same computations, 
on different data, accessing/modifying data on the set which they operate on 
(e.g. fluxes defined on the faces), or, using indirections accessing/modifying 
data defined on other sets (such as coordinate data on connected vertices). 
These indirect accesses are particularly difficult to parallelize when multiple 
threads may try to modify the same data, leading to data races. 

Previous work has utilized one of three approaches for handling data races 
during parallelization~\cite{LULESH:spec,miniaero}: (1) use coloring where the 
iteration set is ``colored'' such that no two iterations of the same color 
modify the same mesh element indirectly, followed by parallel execution of the 
iterations with the same color, (2) use large temporary datasets to stage 
increments without race conditions, and a separate step to gather the 
increments, or (3) use atomics to handle race conditions. However, the amount 
of parallelism, and especially the data locality available to be exploited with 
the above methods have become increasingly limited on modern and emerging 
massively parallel multi-core and many-core architectures. The performance gains 
have been limited particularly on many-core processors such as GPUs with 
thousands of low-power cores, but with modest memory-bandwidth. Thus, reducing 
data movement and exploiting memory locality during execution is vital on such 
devices. On GPUs, the first two techniques, coloring or using temporary 
datasets, end up with poor data locality as one cannot have good data reuse in 
both reading data as well as writing data without conflicts. The third method, 
atomics, are much more expensive operations than regular memory transactions and 
therefore usually lead to low throughput. 

In this paper we explore novel data-movement avoiding and locality exploiting 
algorithms for improving performance of unstructured-mesh applications on GPUs. 
Identifying that the throughput of memory transactions is the main bottleneck, 
we demonstrate how superior execution strategies can be obtained by utilizing 
a combination of techniques from (1) element reordering at thread-block level, 
(2) use of GPU shared memory as an explicitly managed cache and (3) use of 
partitioning algorithms for thread-block formation. We show how these allow us 
to maximize data re-use to the higher-bandwidth shared memory, and optimize 
access patterns to both shared and GPU global memory. More specifically, we make 
the following contributions:
\begin{enumerate}
\item We adopt a caching mechanism on the GPU that loads indirectly accessed 
elements into GPU shared memory. Then use a two-level ``hierarchical coloring'' 
approach to avoid data races, but improve locality over traditional global 
coloring. 

\item We design a reordering algorithm based on graph partitioning that 
increases data reuse within a thread block, also further increasing shared 
memory utilization. 

\item Finally, we apply the above techniques and optimizations to a number of 
representative unstructured-mesh applications to investigate performance on 
modern GPUs, contrasting performance improvements over the state-of-the-art. 
\end{enumerate}

\noindent We demonstrate how the above locality-exploiting algorithms provide 
performance improvements of up to 75\% compared to the state-of-the-art on the 
latest NVIDIA Pascal and Volta GPUs. The algorithms are implemented as an 
open-source software library~\cite{opt-library} which can be used for improving 
performance of existing or new unstructured-mesh applications. 

The rest of the paper is organized as follows: the remainder of Section
\ref{introduction} introduces the basic concepts of unstructured meshes, 
numerical methods based on them and a discussion on related works, Section 
\ref{parallelisation-on-gpu} describes our optimized algorithms and the 
motivation leading to the design. Section \ref{performance} presents the 
performance analysis of the algorithms with experimental results. Finally, in 
Section \ref{conclusion}, we present conclusions from this research.

\subsection{Background}\label{sec:background}
\subsubsection{Unstructured meshes}\label{unstructured-meshes}

\begin{figure}
\centering
\includegraphics[width=4cm]{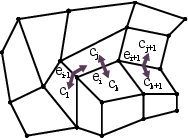}
\caption{Unstructured mesh, the arrow represents the mapping tells $e_i$ is
  connected to $c_j$ and $c_k$.}
\label{fig:unstructured}
\end{figure}

\begin{figure}
\centering
\includegraphics[width=6cm]{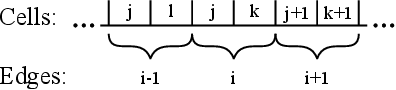}
\caption{A part of the mapping from edges to cells.}
\label{fig:mapping}
\end{figure}

\noindent Unstructured meshes can be abstractly viewed as a collection of sets 
(e.g. nodes, edges, cells, etc.), data defined on these sets (e.g. fluxes, 
coordinates, velocities), and explicit connectivity information between 
sets. The connectivity information, declared as mapping tables are required for 
determining the neighbors of a set element. If we represent sets as consecutive 
indices from zero to the size of the set, then the mapping between two 
sets is represented as an array which stores the index of set elements in the 
second set of the mapping (referred to as the to-set) for every set element of 
the first set (known as the from-set). For the majority of such applications, 
the number of to-set elements connected to each from-set element is fixed (e.g. 
all edges have two vertices). For example consider the mesh illustrated 
in Figure~\ref{fig:unstructured}. Part of the mappings from edges to cells for 
this mesh is detailed in Figure~\ref{fig:mapping}. Given such a mapping, we can 
access the index of those elements that are connected to the current element of 
the from-set from other sets (the to-sets of the mappings). 

The computations on the mesh are declared as a loop over the elements of a set, 
executing some block of computation on each set element (i.e. an elemental kernel), 
while accessing data directly on the iteration set or indirectly through a 
mapping.  If a loop over a set only writes to data defined on that set during the 
elemental kernel, then each iteration of the loop could run in parallel. 
However, for kernels that indirectly increment data, there may be multiple 
from-set iterations that update the same to-set element. Such indirect-loops are 
common in finite volume and finite element applications over 
unstructured-meshes: e.g. when updating state variables in cells using fluxes 
across faces, or when doing matrix assembly. The parallelization of indirect 
loops are non-trivial as the exact elements leading to data races cannot be 
determined from compile time information, given that they are driven by the 
structure of the mesh in general and the mapping tables in particular, which are 
read in during run-time.

Some restrictions that apply is also worth noting here. The first is the use of 
only a single level of mappings. This means that every piece of data that is 
accessed during an iteration over a set is either defined directly on that set, 
or is accessed through at most one level of indirection. However, this 
restriction does not exclude applications using nested indirections, since a 
mapping table can be created to contain the indexes that we access through 
multiple mappings. The second restriction is that the result of the operations 
on the sets are independent from the order of processing the elements of the 
sets (within machine precision). This restriction enables to exploit the 
maximum opportunities for parallelization given that the accuracy of the 
algorithms do not depend on the order of execution. Finally, only  mappings with 
a fixed number of connections (or arity) are considered; such as edges to 
vertices (where the degree is always 2), unlike for a vertices to vertices 
mapping, where this will vary. The natural formalization of most FEM and FV 
algorithm uses mappings with fixed number of connections.

In spite of the above restrictions, the contributions of the research detailed 
in this paper is sufficiently general to be applicable to applications that has 
a computational steps described as an iteration on a set, accessing data on the 
set and indirectly via a mapping (with a fixed arity). Examples include 
assembly, certain types of linear algebra algorithms, flux computations, etc.

\subsection{Related Work}\label{sec:related-works}
\noindent
Algorithms defined on unstructured meshes form an important 
class of applications at many organizations. It is one of the seven dwarfs -- 
common computation-communication patterns or motifs occurring in parallel 
numerical applications -- identified by Colella in 2004~\cite{Colella2004}. 
Discretizations such as finite volumes (FV) or finite elements (FE) often rely 
on these meshes to deliver high-quality results. Indeed there is a large number 
of papers detailing such algorithms, and a wide range of commercial, government, 
and academic research codes (e.g. OpenFOAM~\cite{OpenFoamUserGuide}, Rolls-Royce 
Hydra~\cite{moinier2002edge}, FUN3D~\cite{biedron2017fun3d}). All such 
applications use unstructured meshes in some shape or form, and are often used 
for large experiments, consisting of millions or even billions of mesh elements. 
These codes are generally critical to production and consume large portions of 
high-performance computing systems time. As such, the efficient execution of 
these applications on the parallel architectures of the day has been and 
continues to be crucial to the organizations and stake-holders that have 
invested in them for continued scientific delivery. Over the years, many works 
have discussed and presented techniques for efficient implementations, initially 
focusing on traditional CPU architectures~\cite{mavriplis2002parallel, 
jin1999openmp}, then many-core processors such as GPUs (as we discuss below), 
and even architectures such as FPGAs~\cite{nagy2014accelerating, 
akamine2012reconfigurable}. Many libraries have also been developed targeting 
unstructured-mesh solvers, from classical libraries~\cite{trilinos, PETSc} to 
domain specific languages~\cite{devito2011liszt, giles2012op2, pyfr2016}. 

The adoption of GPUs for these kind of computations has already led to 
considerable speedups over traditional CPU architectures due to the 
massive parallelism available on GPUs~\cite{Reguly2015, ELSEN200810148, 
cohen2009fast}. Other notable works have further looked at improving 
performance. Remacle et al.\ \cite{remacle2016gpu}, explores efficiently solving 
elliptic problems on unstructured hexahedral meshes on GPUs. They use shared 
memory for improving data locality, but advanced techniques, such as reordering 
and partitioning are not utilized. Work done by Castro et al.\ 
\cite{shallow_water} on implementing path-conservative Roe type high-order 
finite volume schemes to simulate shallow flows uses auxiliary accumulators to 
avoid data races while indirectly incrementing. Wu et al.\ 
\cite{wu2013complexity} introduce caching using the shared memory with 
partitioning (clustering), but do not use coloring. Instead they use a 
duplication method similar to that of LULESH and miniAero, as described below. 
Fu et al.\ \cite{fu2014architecting} also create contiguous patches (blocks) in 
the mesh to be loaded into shared memory, although they partition the nodes (the 
to-set) but not the elements (the from-set of the mapping). Furthermore, they do 
not load all data into shared memory, only what is inside the patch. Writing the 
result back to shared memory is done by a binary search for the column index and 
atomic adds, which leads to inefficiencies on the GPU. 

Parallel to the above work, the US Department of Energy labs have released a 
set of proxy applications that represent large internal production codes,
showing some of the computational and algorithmic challenges to be overcome on 
novel and emerging architectures Lulesh \cite{LULESH2:changes}, 
miniAero~\cite{miniaero}, BookLeaf~\cite{bookleaf}, MiniFE~\cite{minife}, 
PENNANT~\cite{pennant}. Out of this suite of codes there are three key 
approaches to handling data races: (1) allocate large temporary arrays where 
the intermediate results (i.e. the increments) are placed, avoiding any race 
conditions, followed by the use of a separate kernel to gather the results, 
(2) use atomics, (3) use coloring. These all lead to increased warp divergence 
and high data access latencies on GPUs; and the use of the temporary array also 
leads to more data allocations and movement, further constraining bandwidth. 

The research detailed in the present work is based on previous work 
in~\cite{op2}, where the OP2 library's GPU parallelization use shared memory on 
GPUs using CUDA for caching with a two level ``hierarchical'' coloring. However, 
we demonstrate superior execution strategies on GPUs with reordering of threads 
and data, to increase data reuse and maximize data locality. Instead of directly 
porting a specific application to use these techniques we present our methods 
as general strategies to accelerate unstructured mesh applications, and in 
particular the indirect increment algorithmic pattern, on GPUs. We have 
created a classical library as open source software~\cite{opt-library} 
incorporating these optimizations. The library can be used for improving 
performance of existing or new unstructured-mesh applications.

Most applications of interest for our work implements finite volume algorithms, 
and low order finite element algorithms, which has a lower computational 
intensity compared to the number of memory transactions. Thus our optimizations 
are targeted to avoid data movement, exploiting locality. In contrast high 
order finite element methods usually have significantly higher computational 
intensity, where there is a higher number of computations per data element 
accessed that can hide the cost of the memory access. While our techniques could 
potentially improve locality, memory bandwidth is less of a concern for such 
applications.

\section{Parallelization on GPUs}\label{parallelisation-on-gpu}

\noindent We begin by outlining the techniques used to effectively optimize 
unstructured mesh applications on GPUs -- some of which are well established 
and commonly used. We briefly show a na\"{i}ve solution, then continue 
with describing various improvements found in the literature, and then present 
our contributions.

\subsection{Traditional parallelization approaches}

\noindent On a GPU, groups of threads (warps) run at the same time, in 
lockstep. As such it is not efficient to execute computations of different 
length on different threads. Consequently, the usual practice is for each thread 
to take responsibility for the computation on one element of the set. This can 
be viewed as running one iteration per thread in the loop over a given set, also 
known as the iteration set or from-set. This allows the number of computations 
to be fixed, where the amount of data involved is fixed in the dimension of the 
mapping and the data arrays.

\begin{figure}[Htpb]
  \centering
  \includegraphics{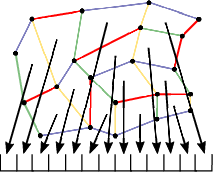}
  \caption{Schematic figure of the global coloring approach. In each kernel
  launch, the kernels work on edges of the same color. The arrows
  represent the individual pieces of data loaded indirectly when executing the 
color red.}
  \label{fig:unstructured_global}
\end{figure}

As mentioned before, care must be taken when writing parallel code to avoid 
data races when different threads modify the same data. There are three 
approaches discussed in the literature. The first is to color each thread 
according to the indirect data it writes, so that no two threads with the same 
color write the same data, and enforce ordering between colors using 
synchronization~\cite{Zegard2013}. On the GPU, one would do multiple kernel 
launches corresponding to the colors, so there is no concurrent writes between 
threads in the same kernel. We call this the \emph{global coloring} approach 
(Figure \ref{fig:unstructured_global}). The disadvantage here is that there is 
virtually no data reuse: when multiple elements write the same data, they are 
scheduled for execution in different launches. Since these operations also tend 
to read data through the same mappings, there is no data reuse in the reads 
either. Compounding the issue is low cache line utilization where elements of 
the same color are not neighbors in the mesh, and therefore unlikely to be 
stored in consecutive memory locations.

The second approach is to serialize the indirect updates by means of locks or
atomic additions \cite{Kraus:2014:ACC:2691158.2691164}. This is considerably 
expensive on the GPU, since the whole warp has to wait at the synchronization 
step leading to warp divergence.

The third solution is the use of a large temporary array that stores the
results for each thread separately, avoiding race conditions by formulation 
\cite{LULESH:spec,miniaero}. However, after the computation finishes, a 
further kernel is required to gather the results corresponding to one data 
point. This suffers from the problem of not knowing how many values one thread 
has to gather, and as a result warps could diverge significantly, and 
memory access patterns are less than ideal. They can be good either for the 
write or the read, but not for both. Also, the size of the temporary array is the 
number of elements multiplied by the dimension of the mapping. As a result, it 
can be large, for example, in LULESH, it is \(8 \times 3 \times numElem \) 
in our measurements (where $numElem$ is the size of the from-set in LULESH), 
compared to the array defined on nodes where these values will ultimately end 
up, which is roughly the same as the number of elements themselves.

\subsubsection{Array-of-Structures (AoS) vs Structure-of-Arrays (SoA)} 
\label{aos-to-soa}

\noindent Due to the lockstep execution, consecutive threads in a warp read  
memory at the same time. Therefore, the layout of the data in the memory is an 
important factor for performance. There are two commonly used 
layouts~\cite{sharma2015data}: (1) Array-of-Structures (AoS) layout, where the
data associated with one element is in consecutive places in the array (and thus
in memory) and (2) Structure-of-Arrays (SoA) where the components of elements
are stored consecutively e.g. the first data component of the elements are in
the beginning of the array followed by the second, etc.

Although in most cases the SoA gives better performance on GPUs and better
vectorization on CPUs, the AoS layout is still commonly used on CPU
architectures with large caches. In the case of the AoS layout, consecutive
threads read data from strided addresses in memory and thus more cache lines 
are required to satisfy one transaction. This would be compensated by 
subsequently reading the other components, but may have a negative effect on 
 GPUs due to their small caches. Conversely, with the SoA layout, the 
threads read data next to each other, which means that the data needed by 
consecutive threads are most probably in the same cache line resulting in 
coalesced memory transactions. However, when indirections are involved, these
access patterns become more complicated --- even with the SoA pattern,
consecutive threads may not be reading consecutive values in memory, and
therefore cache line utilization degrades. The choice of data layout in
unstructured mesh computations is therefore highly non-trivial, as we show
later.

\subsection{Shared memory approach}\label{shared-memory-approach}

\noindent Considering the three data race avoiding approaches, we see that they 
all only make use of the GPU global memory. As such one technique to further 
improve performance is by reducing memory accesses to the GPU global memory. To 
this end, the OP2 library~\cite{op2} targets the use of the shared memory on 
the GPUs. Shared memory is only shared within thread blocks, but has much lower 
access latency and higher bandwidth than the global memory. The idea is to 
collect the data required to perform the computations and load it into shared 
memory. Then, during computation, the indirect accesses are to the shared
memory, and the result can also be stored there. After computations by all 
threads in the block have completed, the contents of the shared memory can be 
written back to global memory. One immediate advantage of this approach is that 
the fetching and writing back of data from/to the global memory can be done by 
the threads independently of the actual threads that will be carrying out the 
computations on them. Particularly, reading/writing can be done in the order 
in which the data is laid out in memory, ensuring maximum utilization of 
cache lines. With the AoS layout, data can be read in contiguous chunks as 
large as the number of components in the structure.

\begin{figure}[Htpb]
  \centering
  \includegraphics[width=0.5\textwidth]{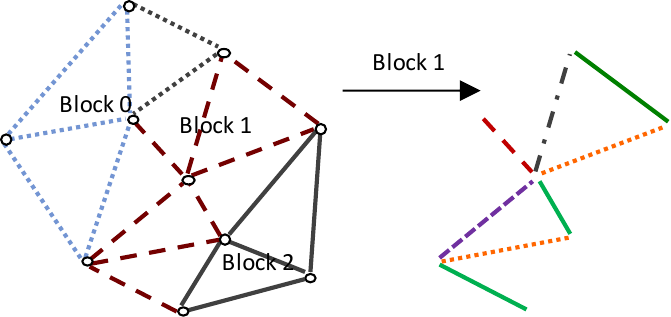}
  \caption{Schematic figure of the two levels of coloring.}
  \label{fig:multilevel}
\end{figure}

\begin{figure}[Htpb]
  \centering
  \includegraphics{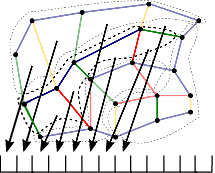}
  \caption{Schematic figure of the hierarchical coloring approach. The thread
  blocks are circled with dashed lines. The arrows represent the individual data
  points loaded.}
  \label{fig:unstructured_hier}
\end{figure}

The use of shared memory of course leads to one additional complication. 
Writing back the updated patches of shared-memory to GPU global memory may lead 
to data races. This leads to the use of a \emph{two-layered coloring} or 
\emph{hierarchical coloring}~\cite{op2} scheme. The two levels of coloring are 
illustrated in Figure \ref{fig:multilevel}, and the associated data accesses are 
shown in Figure \ref{fig:unstructured_hier}. The first level of coloring is to 
avoid data races when thread blocks write the result back to global memory and 
the second level is to avoid threads writing their results into shared memory at 
the same time. 

Algorithm~\ref{code:shared} details the steps carried out by a CUDA kernel for
executing within this two-level coloring scheme. All indirect data accessed by
the block (which is identified during a preprocessing phase) is fetched from
global to shared memory to shared memory. As noted in Section~\ref{aos-to-soa},
data is usually composed of data points with multiple components (for example,
x-y-z coordinates), so this operation consists of two nested loops: (1) an
iteration over the data points, and within that, (2) an iteration over the data
corresponding to the data point. For the SoA layout, only the outer loop needs
to be parallelized, as this will cause parallel read operations to access memory
addresses next to each other. For the same reason, if the AoS layout is used,
both parallel loops need to be parallelized (i.e.  collapsed into one). The data
layout in shared memory is best be set to SoA: our measurements showed a
consistent degradation in performance when switching to AoS layout, due to the
spatial locality described in Section \ref{aos-to-soa}: it leads to fewer bank
conflicts.

After the data is loaded into shared memory, each thread executes the main body 
of the kernel, and outputs are placed into registers.  Next, the threads update 
the result in shared memory with their increments. Finally the updated data is 
written back to global memory.

\begin{algorithm}
  \begin{algorithmic}
    \State \lstinline!tid = blockIdx.x * blockDim.x + threadIdx.x!
    \State \lstinline!bid = blockIdx.x!
    \ForAll {\lstinline!data_point! $\in$ \lstinline!indirect_data!}
      \ForAll {\lstinline!d! $\in$ \lstinline!data_point!}
        \State \lstinline!shared[shared_ind(d)] = global_indirect[d]!
      \EndFor
    \EndFor
    \State \lstinline!__syncthreads()!
    \State \lstinline!result = computation(shared[mapping[tid]], global_direct[tid])!
    \State \lstinline!__syncthreads()!
    \State fill shared memory with zeros
    \State \lstinline!__syncthreads()!
    \For {\lstinline!c = 1! $\ldots$ \lstinline!num_thread_colours!}
      \If{\lstinline!c == thread_colours[tid]!}
        \State increment shared with result
      \EndIf
      \State \lstinline!__syncthreads()!
    \EndFor
    \ForAll {\lstinline!data_point! $\in$ \lstinline!indirect_data!}
      \ForAll {\lstinline!d! $\in$ \lstinline!data_point!}
        \State increment \lstinline!global_indirect_out! with shared
      \EndFor
    \EndFor
  \end{algorithmic}
  \caption{Algorithm to use the shared memory to preload indirect data accessed
  within a thread block. \lstinline!global_indirect! holds the data indirectly
  read, \lstinline!global_indirect_out! holds the result of the iteration.}
  \label{code:shared}
\end{algorithm}

One other benefit from using shared memory with hierarchical coloring is the 
improved data reuse within the block. Each piece of data has to be loaded 
from global memory only once, but can be used by multiple threads (e.g. data on 
a shared edge between two triangles). However, the greater the reuse, the more 
thread colors we have: the number of colors is no less than the number of 
threads writing the same data. Since the number of synchronizations also grows 
with the number of thread colors (more precisely, it is the number of colors 
plus two, one before and one after the computation if the input and the 
increment are stored separately in shared memory), there is a trade-off between 
the number of synchronizations and data reuse. Our measurements showed that if 
the kernel is memory-bound, the greater data reuse leads to increased 
performance, but the trade-off is non-trivial, as we will demonstrate in 
Section~\ref{performance}.

\subsection{Increasing data reuse}\label{increasing-data-reuse}
\noindent Building on the shared-memory with hierarchical coloring, the first 
contribution of our work attempts to further increase data reuse through 
reordering of elements. Specifically, reordering of the elements in the
from-set (which map directly to the threads), allows us to control 
how CUDA thread blocks are formed and how much data reuse can be achieved. 
With the shared-memory approach, the benefit of data reuse is twofold: it 
decreases the number of global memory transactions and decreases the size of 
shared memory needed, which leads to greater occupancy. Two different 
approaches to re-ordering is explored (1) the sparse matrix bandwidth reducing 
Gibbs-Poole-Stockmeyer algorithm~\cite{gps} and (2) graph partitioning.

\subsubsection{Gibbs-Poole-Stockmeyer-based reordering}

\begin{figure}%
  \centering%
  \subfloat[][]{%
    \centering%
    \includegraphics[width=5cm]{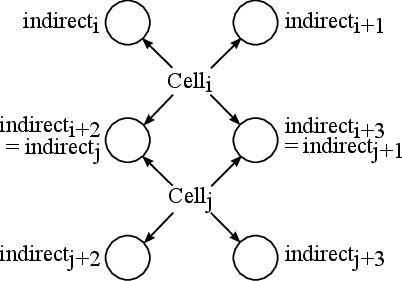}%
    \label{fig:mesh2grapha}%
    }%
  \qquad
  \subfloat[][]{%
    \centering%
    \includegraphics[width=5cm]{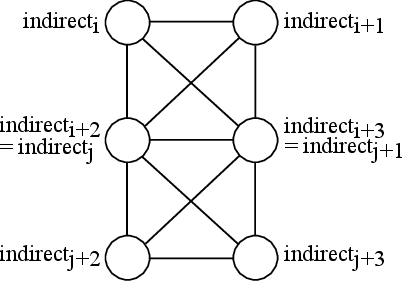}%
    \label{fig:mesh2graphb}%
    }%
  \caption[]{An example of converting a mesh (shown in \subref{fig:mesh2grapha},
  with mapping dimension 4) to a graph (on Figure \subref{fig:mesh2graphb}) for
  the GPS algorithm.}%
  \label{fig:mesh2graph}
\end{figure}

\noindent For serial implementations of computations on graphs (typically on 
CPUs), the Gibbs-Poole-Stockmeyer algorithm (GPS,~\cite{gps}) is a heuristic 
algorithm that increases spatial and temporal locality when traversing the 
nodes. For example, considering a mesh with edges and nodes, where the edges are 
the elements of the from-set of the mapping, and the nodes form the to-set, GPS 
would renumber the nodes and change the order of traversal. The renumbering is 
done by going through the nodes in a breadth-first manner from two distant 
starting points, and then renumbers the nodes so that the levels of the 
resulting spanning trees will constitute contiguous blocks in the new 
permutation. After renumbering its points, which by design improves spatial 
locality, we order the edges of the graph lexicographically, so that consecutive 
threads (or spatial iterations in serial implementations) have a higher chance 
of accessing the same points, which improves temporal locality (data reuse). 
The algorithm can be generalized to meshes by transforming each element into a
fully connected graph of its points and then taking the union of these. An
example of this is shown on Figure \ref{fig:mesh2graph}.

There are several straightforward generalizations to handle multiple sets and
mappings (e.g. vertices, edges, cells and their connections).  The first is to
assume that all the mappings describe a similar topology, so the elements can be
reordered based on only one of the mappings (as described above), then reorder 
the points accessed through the other mappings by, for example, a greedy method.
Another approach could be to reorder every data set separately, and then reorder
the elements based on the new order of the accessed points, combining the
separate data sets (and corresponding mappings) in some way. Since the mappings
in the applications we measured are very similar topologically (in fact, except
for one of the applications we tested, Airfoil, there is only one mapping in 
each application), we used the first method. However, the algorithm fails to 
take into account that on the GPU the threads are grouped into blocks, and data 
reuse can only realistically be exploited within blocks. This results in blocks 
that are ``pencil-shaped''. The next algorithm addresses this limitation.

\subsubsection{Partitioning based reordering}

\begin{figure}[Htpb]
  \centering
  \includegraphics{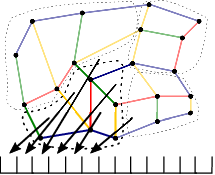}
  \caption{Schematic figure of the hierarchical coloring approach with
  partitioning. The thread blocks are circled with dashed lines. The arrows
  represent the individual pieces of data loaded; note that this is less than
  in Figure \ref{fig:unstructured_hier}.}
  \label{fig:unstructured_part}
\end{figure}

\noindent To increase data reuse within a block is equivalent to decreasing 
data shared \emph{between} the blocks, more specifically, to decrease the 
number of times the same data is loaded in different blocks (see 
Figure~\ref{fig:unstructured_part}). With the shared memory approach, data needs 
to be loaded only once per block. So the task is to partition the elements into 
blocks of approximately the same size in such a way that when these blocks are 
assigned to CUDA thread blocks, the common data used (loaded into shared 
memory) by different blocks is minimized.

Let $G_M$ be a graph constructed from the original mapping, where the points are
the threads, and there is an edge between them if and only if they access the
same data, and let $P_{G_M} = \{B_1, \ldots, B_n\}$ be a partition of this graph
with $n$ blocks. This works even with multiple mappings. If there is a set of 
blocks $B_{d_1}, \ldots, B_{d_k}$ that access the same piece of data, then they 
form a clique in $G_M$ in the sense that between any pair of blocks $B_{d_i}$ 
and $B_{d_j}$ (where $1 \le i,j \le k$), there is an edge of $G_M$ between $u$ 
and $v$ such that $u \in B_{d_i} \wedge v \in B_{d_j}$. Note that the cliques 
have $0.5 \cdot (k^2 - k)$ edges, which is a monotone increasing function in 
$k$, since $k \ge 1$ (there is at least one block writing each data point, 
otherwise it is of no relevance). That means that partitioning using the usual 
objective of minimizing the number of edges between blocks is a good heuristic 
for maximizing data reuse within the blocks.

We chose the k-way recursive partitioning algorithm used by the 
METIS~\cite{metis} library to partition the graph $G_M$. It is a hierarchical
partitioning algorithm, where it first coarsens the graph by collapsing nodes, 
then partitions using the recursive bisection algorithm, and then finally while 
progressively un-coarsening the graph, locally optimizes the cuts. The algorithm 
attempts to maintain equal block sizes in the resulting partition, however, it 
is not always possible because of the underlying algorithm. Since CUDA launches 
thread blocks with equal size, this must be the maximum of the block sizes in 
the created partition. Consequently some threads do not do any work, 
lowering occupancy. One of the tuning parameters for the algorithm is the the 
load imbalance factor, which can be used to specify the tolerance for this 
difference. It is called load imbalance because METIS was originally used 
for distributing computation, ie. load, in a distributed memory system. The 
Load imbalance factor is defined as $l = n\max_j \left\{\mathrm{size}(B_j) 
\right\}$, where $n$ is the number of blocks and $\mathrm{size}(B_j)$ is the 
size of the $j$th block. Due to the local optimization in the un-coarsening 
phase, it is impractical to set this parameter to $1$ (meaning the block sizes 
must be exactly the same). We found that a tolerance of $1.001$ works well in 
practice for our needs.

We design the block size to be a tuning parameter, which specifies the actual
block size of the launched GPU kernels. The number of working threads
cannot exceed the block size. To account for this in partitioning, we calculate 
a new block size ($S'$) and tolerance ($l'$) with margins for the imbalance:
\begin{align}
  S' &= \left\lfloor \frac{S}{l} \right\rfloor \\
  l' &= \frac{S + \epsilon}{S'},
\end{align}
where $S$ is the original block size, $l$ is the original load imbalance
parameter and $\epsilon$ is an empirical tuning parameter to create as large
blocks (within the limit) as possible. This support for a variable number of 
working threads (ie. to determine if the current thread should do any actual 
computation) also incurs a slight overhead of having to load the start and end 
index for each block. We found this overhead to be minimal in practice.

Due to the way loads and stores work on the GPU, what actually affects
performance is not the number of data points accessed, but rather the number of
cache lines (of size 32 bytes on the hardware used) that are accessed. A simple 
heuristic reordering of data points is used to account for this. 

The idea is to group data points together (in a contiguous chunk of memory)
that are read/written by the same set of blocks: this makes them more likely to
be loaded in the same cache line. This is even more important when more blocks
access the same group of data points (set elements on the boundary), since then
inefficiencies will worsen performance for each of these blocks. As a simple
heuristic, we group data points with the same number of blocks that access them
together (by sorting), and within these groups, we sort by the indices of the
accessing blocks lexicographically.

It must be noted that GPS and METIS were developed for distributing workloads on
computing clusters that typically have much larger block size to total size
ratio. This also caused the reordering (partitioning) phase, running on a 
single CPU core, to be quite long: for example, the GPS reordering took 5 
seconds, the partitioning 60--90 seconds and the hierarchical coloring 5--9 
seconds (depending on the block size and whether the mesh was partitioned) on 
the mesh used by Airfoil. However, this is a one-off cost: the reordering can 
be reused many times later. Further improvement could be achieved by using the 
parallel versions of the METIS library: ParMETIS\cite{parmetis} and the alpha 
version mt-METIS\cite{mtmetis} libraries. Partitioning algorithms targeting 
specifically small partition sizes required for CUDA thread blocks, and 
improving its performance were out of scope for this work.

\subsection{Further optimizations}\label{optimisations}

\noindent There are a number of further optimizations we introduced to improve
performance. We can increase the number of loads or stores in flight 
to/from global memory by using CUDA's built-in vector types 
(\lstinline!float2!, \lstinline!float4! and \lstinline!double2!). This way, 
each thread will load multiple consecutive values from memory during a single 
transaction. This is useful to increase the efficiently of loads that are 
already coalesced.

When updating the shared memory with increments, the threads within a block can 
be sorted by their color. Then threads with the same color will be next to each 
other, so warps will have fewer threads of different colors. This results in 
reduced warp divergence on average.

Marking pointers to data on the GPU with \lstinline!__restrict__! and 
\lstinline!const! where applicable enables the compiler to apply further 
reordering optimizations, which it would not have deemed safe to do otherwise. 
\lstinline!__restrict__! instructs the compiler that the pointers do not 
\emph{alias} one another, ie. do not point to the same memory space. The 
\lstinline!const! enables the compiler to place the data in texture cache that 
has lower latency than the global memory.

\section{Performance}\label{performance}

\noindent For the remainder of the paper, we present a detailed investigation 
into the performance implications of the locality-exploiting optimizations on a 
number of representative unstructured-mesh applications. The applications are 
all well established in literature and are representative of several domains 
that make use of unstructured mesh codes. Our aim is to present the performance 
of the state-of-the-art on GPUs with these applications and then contrast the 
performance gained with our contributions. 

\subsection{Experimental setup}\label{experimental-setup}

\noindent The GPU systems used in performance analysis is detailed in 
Table~\ref{tab:GPU_datasheet}. These consist of two of NVIDIAs latest 
high-performance computing Tesla GPUs, namely the P100 and V100. The GPUs are 
hosted in a server with an Intel Xeon CPU E5-1660 (3.20GHz base frequency, 1 
socket with 8 cores) running Ubuntu 16.04. The nvcc compiler with CUDA version 
9.0 (V9.0.176) is used. 
\begin{table}
\centering
\begin{tabular}{lcc}
\hline
  & P100 (GP100) & V100 (GV100)\\ \hline\hline
  Streaming Multiprocessors (SM) 		& 56	& 80\\ \hline
  Max Thread Block Size				& 1024	& 1024 \\ \hline
  Max Warps / Multiprocessor 			& 64 & 64	\\ \hline
  Max Threads / Multiprocessor		& 2048 & 2048	\\ \hline
  Max Thread Blocks / Multiprocessor 	& 32 & 32	\\ \hline
  Max Registers / Thread& 255 & 255	\\ \hline
  Shared Memory Size / SM	& 64 KB & 96 KB	\\ \hline
  Max 32 - bit Registers / SM			& 65536 & 65536\\ \hline
  Memory Size							& 16 GB	& 16 
GB\\ \hline
  L2 Cache Size						& 4096 KB & 6144 KB\\ 
\hline
  Peak Stream Bandwidth					& 495 GB/s & 742 GB/s\\ 
\hline
\end{tabular}
  \caption{Important informations about the NVIDIA Tesla P100 and V100 GPUs
  \cite{Pascal_whitepaper, Volta_whitepaper}}
\label{tab:GPU_datasheet}
\end{table}

Both runtime performance as well as low-level metrics on the GPUs such as 
achieved bandwidth and occupancy is utilized to understand the bottlenecks 
affecting performance. There are three key factors of interest: (1) 
the coloring approach (global or hierarchical) giving the best performance, 
(2) method of data reordering (no-reordering, GPS-based, or graph 
partitioning-based) and (3) data layout (AoS or SoA). All combinations of these 
are evaluated and compared to each other and a state-of-the-art reference 
implementation.

When comparing performance of different versions, we use the achieved bandwidth
as the key performance metric. Emphasis on bandwidth is justified given that 
all of the test applications, as we demonstrate, are memory-bound. Achieved 
bandwidth is calculated by the formula: $$\frac{\sum_{d} w_dS_d}{T} \cdot I,$$ 
where $d$ iterates over the datasets, $w_d$ is $2$ if the data is read and 
written, $1$ otherwise, $S_d$ is the size of the dataset (in bytes), $T$ is the 
overall runtime of the kernel and $I$ is the number of iterations. A number 
of additional metrics were also collected, these include:
\begin{itemize}
\item data reuse factor (the average number of times an indirectly accessed
data point is accessed),
\item the number of read/write transactions from/to global memory, which is
closely related to the data reuse factor but is affected by memory access
patterns, and therefore cache line utilization,
\item the occupancy reflecting the number of threads resident on the SM versus
the maximum - the higher this is, the better chance of hiding the latency of
compute/memory operations and synchronization
\item the percentage of stalls occurring because of data requests, execution
dependencies, or synchronization,
\item the number of block colors; the higher it is, the less work in a
single kernel launch, which tends to lead to lower utilization of the GPU,
\item the number of thread colors; the higher this is the more
synchronizations are required to apply the increments in shared memory ---
but also strongly correlates with data reuse,
\item warp execution efficiency (ratio of the average active threads per warp
to the maximum number of threads per warp).
\end{itemize}
Studying runtime performance and the above metrics enables us to understand 
and explain why certain variants are better than others.

\subsubsection{Airfoil}

\noindent Airfoil, implemented using the OP2 DSL~\cite{op2} is the smallest, 
best understood and most thoroughly studied among the applications we explored. 
It is representative of large industrial Finite Volume CFD applications and 
implements a non-linear 2D inviscid airfoil code using an unstructured grid. A 
finite-volume discretization is used to solve the 2D Euler equations with a 
scalar numerical dissipation. The algorithm iterates towards the steady state 
solution, in each iteration using a control volume approach, meaning the change 
in the mass of a cell is equal to the net flux along the four edges of the cell, 
which requires indirect connections between cells and edges. Two versions of the 
code exists, one implemented with OP2's C/C++ API and the other using OP2's 
Fortran API~\cite{giles2012op2,op2-repo}.

The application consists of five parallel loops in total: \texttt{save\_soln}, 
\texttt{adt\_calc}, \texttt{res\_calc}, \texttt{bres\_calc} and \texttt{update}.
Here we focus on \texttt{res\_calc}, as it has indirect increments and about 
70\% of the total runtime of the application is spent in this parallel loop on 
GPUs when using a global coloring. The loop contains both indirect reads and 
writes. It iterates through edges (i.e. the from-set), and computes the flux 
through edges using data accessed indirectly on the two cells adjacent to each 
edge. The \texttt{res\_calc} loop is called 2000 times during the  
execution of the application and performs about 100 floating-point operations 
per mesh edge. In each iteration, it reads 5 and increments 4 double values from 
each of the 2 indirectly accessed cells, and reads 2 double values from each of 
the 2 indirectly accessed nodes. 

Table \ref{tab:airfoil_counters_glob} show the effect of various optimizations 
on the Airfoil application's \texttt{res\_calc} kernel, during the execution on 
a mesh with 2.8 million cells.

\begin{table}[Htbp]
\centering
\resizebox{\columnwidth}{!}{
\begin{tabular}{|R{3cm}|cc|c|c||cc|c|c||c|}\hline
Coloring  & \multicolumn{4}{c||}{Global} & \multicolumn{4}{c||}{Hierarchical} & 
\begin{tabular}{@{}c@{}}Original \\ Hierarchical\end{tabular}\\\hline
Reordering & \multicolumn{2}{c|}{none} & GPS & partition & 
\multicolumn{2}{c|}{none} & \multicolumn{2}{c||}{partition} & none\\\hline
Data layout &    AOS      &   SOA     &     SOA      &    SOA & AOS      &   SOA 
    &     AOS      &    SOA & SOA\\\hline
Bandwidth (GB/s)&  $72$ & $94$ &  $106$&  $66$&    $ 211$ &   $ 215$ &
$ 228$ &   $ 239$ & $ 233$\\\hline
Runtime (ms) & $6.12$ & $4.65$ & $4.15$ & $6.64$ & $2.07$ & $2.03$ & $1.92$ & 
$1.83$ & $1.91$\\\hline
Achieved Occupancy&  $0.63$ &  $0.45$ &      $0.45$&            $0.45$&    
$0.44$ &   $0.43$ &      $0.44$ &   $0.43$ & $0.42$\\\hline
Global Memory Read Transactions&  \num{52424}k & \num{45781}k & \num{41246}k& 
\num{66775}k& \num{21142}k & \num{21275}k & \num{13885}k & \num{14325}k& 
\num{21866}k\\\hline
Global Memory Write Transactions& \num{14007}k & \num{14737}k & \num{13773}k& 
\num{20733}k& \num{5807}k  & \num{5871}k & \num{3429}k  & \num{3628}k & 
\num{6384}k\\\hline
Number of (Block) Colours&  $5$&  $5$ &      $5$ &            $7$&    $       4$ 
&   $       4$ &      $       8$ &   $       8$ &   $       5$\\\hline
Number of Thread Colours&-&-&-&-&    $       3$ &   $       3$ &      $       4$ 
&   $       4$ &   $       3$\\\hline
Reuse Factor&-&-&-&-&    $       2$ &   $       2$ &      $     3.6$ &   $     
3.6$ &   $       2$\\\hline
Issue Stall Reasons (Synchronization)&-&-&-&-&    $  11\%$ &   $   10\%$ &      $ 
 15\%$ &   $  14\%$ &   $  14\%$\\\hline
Issue Stall Reasons (Data Request)&-&-&-&-&    $  69\%$ &   $  70\%$ &      $  
62\%$ &   $  64\%$ &   $  55\%$\\\hline
Block Size &\multicolumn{8}{c||}{$480$} &   $128$\\\hline
\end{tabular}
}
\caption{Low-level performance metrics of Airfoil's \texttt{res\_calc} kernel - 
global coloring vs hierarchical coloring (2.8 million mesh cells). The last 
column details the measured performance of the original code.}
\label{tab:airfoil_counters_glob}
\end{table}

\emph{Global coloring}: We see that using the SoA layout improves performance. 
As discussed in Section~\ref{aos-to-soa}, with SoA threads in a warp access data 
addresses that are near each other. The improvement can also be seen in 
the number of global memory read transactions as it is roughly $87\%$ of that 
with AoS layout. Adding the GPS renumbering improves performance 
further by placing data points that are accessed in consecutive threads close 
to each other. Now there is a $19\%$ reduction in global read transactions 
compared to the baseline AoS. Given that the partition based reordering is 
primarily intended for the hierarchical coloring, it does not improve on the 
Global coloring. The reason being that partitioning groups threads 
that access the same data together, while the global coloring puts them into 
different kernel launches, eliminating any chance for spatial reuse.

\begin{figure}[Htbp]
\centering
\includegraphics[width=10cm]{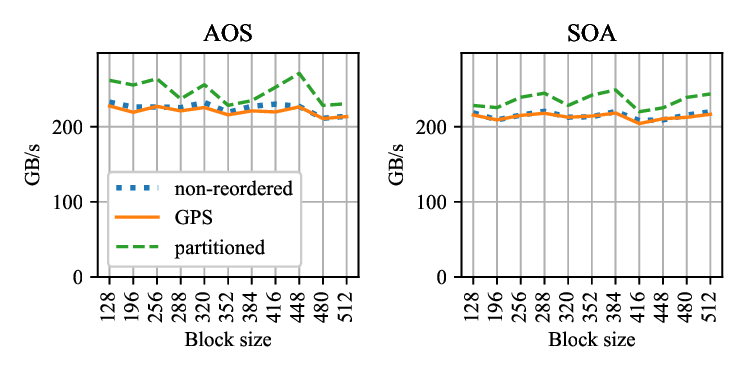}
\caption{Airfoil's \texttt{res\_calc} bandwidth on a dataset with $2880000$ 
cells with hierarchical coloring.}
\label{fig:airfoil_bw-vs-bs_hier_large}
\end{figure}

\emph{Hierarchical coloring:}  The key goal of this strategy is to better 
exploit data reuse by using the GPU shared memory. The effectiveness of the 
approach show immediately due to the significant reduction in the number of 
global transactions in Table~\ref{tab:airfoil_counters_glob}. At block size 
$480$, there is roughly a $60\%$ decrease in global read and write 
transactions, leading to three times the performance. Throughput for different 
block sizes is shown in Figure~\ref{fig:airfoil_bw-vs-bs_hier_large}.

We also see that reordering using partitioning is indeed more effective. With
a block size of 448, data reuse increases from $2$ on the reference version,
to $3.6$, leading to the $19\%$ performance gain over the version without
reordering (AoS layout). This is also consistent with the number of global
transactions: there is a $35\%$ decrease in the number of reads and $41\%$
decrease in the number of writes, and a decrease in the percentage of stalls
occurring because of data requests: $61\%$ with partitioning, $68\%$ without.

\begin{figure}[Htbp]
\centering
\includegraphics[width=10cm]{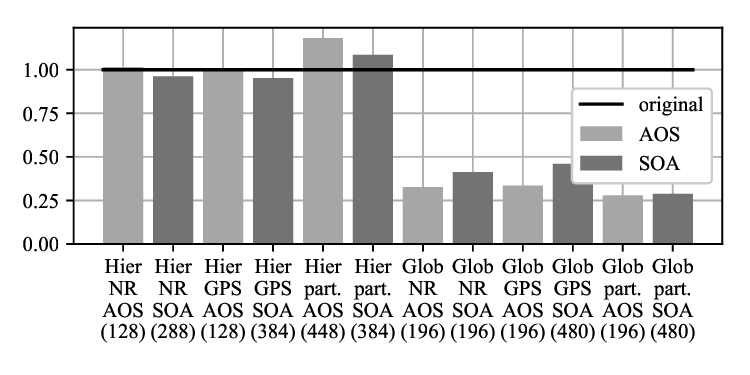}
\caption{Airfoil's \texttt{res\_calc} kernel speedup compared to the original 
code (on a mesh with $2800000$ cells. The block sizes are shown in parentheses, 
the reordering algorithms are: the original reordering (NR), GPS reordering and 
partitioning (part.)}
\label{fig:airfoil_speedup_large}
\end{figure}

With the increased reuse, the number of thread colors is also larger ($4$
versus $2.2$) and this leads to more synchronization. With reordering, $14\%$
of the stalls were caused by synchronization, up from $9\%$. This is further 
illustrated by Figure \ref{fig:airfoil_speedup_large} that shows the relative 
speedup compared to the original OP2 version (its low-level metrics are detailed 
in the final column of Table~\ref{tab:airfoil_counters_glob}). In this case, the 
original version also used the shared memory approach, so the performance gains 
are caused by the reordering. In the original version (hierarchical coloring, SOA
layout) $56\%$ of the total time is spent in \texttt{res\_calc}. The best original
version used the SOA data layout, with reordering we achieved $19\%$ speedup on
\texttt{res\_calc} with AOS layout.
However with AOS layout we lose performance in direct kernels, therefore regarding
the whole application one can reach better performance with the SOA layout.
The best performing setting of airfoil reached about $3.3\%$ speedup on the whole 
application. The useful bandwidth of the best performing version of 
\texttt{res\_calc} (our implementation) reached $55\%$ of the peak stream bandwidth
of the P100 GPU.

\begin{figure}[Htbp]
\centering
\includegraphics[width=10cm]{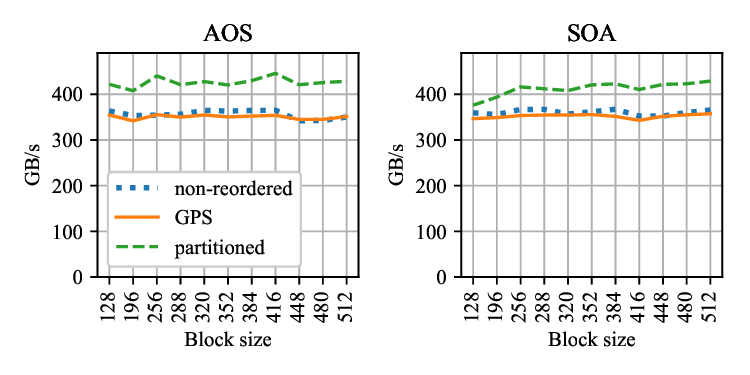}
\caption{Airfoil's \texttt{res\_calc} bandwidth on a dataset with $2880000$ 
cells with hierarchical colouring, on Volta architecture.} 
\label{fig:airfoil_bw-vs-bs_hier_large_volta}
\end{figure}

Similar results were obtained on the newer Volta GPUs (V100) as illustrated in 
Figure~\ref{fig:airfoil_bw-vs-bs_hier_large_volta}). The absolute value of the
bandwidths are (understandably) higher. On the V100, AOS achieved $28\%$ 
speedup in the kernel compared to the original code with $446$ GB/s bandwidth 
($60\%$ of the peak bandwidth), but the it still lacks behind SOA regarding the
whole application. The best SOA version achieved $21\%$ speedup. Since 
\texttt{res\_calc} takes around $54\%$ of the total run time this speedup lead 
to $11\%$ performance increase on the whole application.

\subsubsection{Volna}
\noindent The next application we explore here, Volna, is in fact a 
production/research code for shallow water simulation capable of handling the 
complete life-cycle of a tsunami (generation, propagation and run-up along the 
coast)~\cite{dutykh2011volna}. The simulation algorithm works on unstructured
triangular meshes and uses the finite volume method. Volna is written in C/C++
and is converted to use the OP2 library~\cite{op2-volna2018}. Volna spends 
most time in three kernels: \texttt{computeFluxes}, \texttt{SpaceDiscretization} 
and \texttt{NumericalFluxes}. Out of these, we focus on the 
\texttt{SpaceDiscretization} kernel that iterates on edges accessing data 
indirectly on cells, $60\%$ of total execution time is spent in this kernel. In 
each iteration, \texttt{SpaceDiscretization} reads 1 and increments 4 float 
values from each of the 2 indirectly accessed cells, and reads 7 float and 1 
integer values directly. A notable difference in Volna is that the execution 
with single precision is adequate for solution accuracy. As such we benchmark it 
with single precision floating-point mathematics, on a mesh containing 2.4 
million triangular cells, simulating a tsunami run-up to the US pacific coast. 

\begin{figure}[Htbp]
  \centering
  \includegraphics[width=10cm]{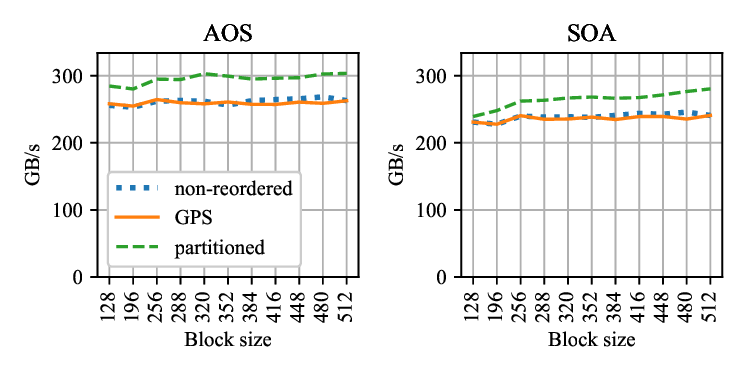}
  \caption{Volna's \texttt{SpaceDiscretization} kernel bandwidth on a mesh with
  $3589735$ edges using hierarchical coloring.}
  \label{fig:volna_bw-vs-bs_hier}
\end{figure}

Figure~\ref{fig:volna_bw-vs-bs_hier} and Table~\ref{tab:volna_counters_hier} 
details the performance metrics observed for the \texttt{SpaceDiscretization} 
kernel in Volna. We concentrate solely on the hierarchical coloring variants 
given their superior performance to global coloring. The reordering by 
partitioning again improves performance. It increases reuse from $1.5$ to 
$2.8$ and decreases the number of global transactions by $18\%$ for reads and 
$37\%$ for writes. The larger reduction in writes can be explained by the fact 
that the calculation only reads data defined on the iteration set directly.

Again, recall that the AoS version uses adjacent threads to load adjacent 
components of data points. Additionally, given the use of single precision 
values for Volna, one thread loads $4$ single precision values into shared 
memory using the built-in vector type \lstinline!float4!. Consequently, more 
data is transferred at the same time, providing a $2\%$ and $4\%$ reduction
in global memory transfers for reads and writes, respectively. This leads 
to performance improvements of $292\,\text{GB/s}$ versus $268\,\text{GB/s}$.

Low register counts ($28$--$32$) and single-precision data types also 
resulted in achieving a higher occupancy on the GPU compared to Airfoil. This, 
we believe explains why performance appears to be independent of the block size 
as shown in the Figure \ref{fig:volna_bw-vs-bs_hier}. 

We also see that using partitioning does not increase the number of thread 
colors significantly, as we observed on Airfoil. The increase of colors are 
from $3$ to $4$. As such the overall synchronization overhead is also smaller. 
The percentage of stalls caused by synchronization increases from $12\%$ to just 
$15\%$. Of course, with high occupancy, the latency caused by synchronization 
can be better hidden by running warps from other blocks. 

\begin{figure}[Htbp]
\centering
\includegraphics[width=10cm]{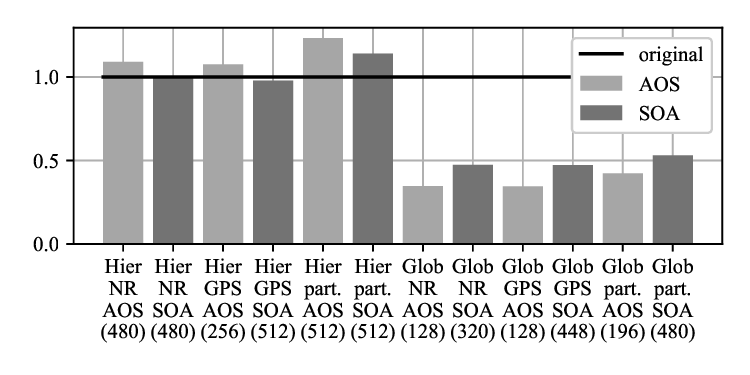}
\caption{Volna's \texttt{SpaceDiscretization} kernel speedup compared to the 
original code, done on a mesh with $3589735$ edges. The block sizes are shown 
in parentheses, the reordering algorithms are: the original reordering (NR), 
GPS reordering and partitioning (part.)}
\label{fig:volna_speedup}
\end{figure}

As can be seen in Figure \ref{fig:volna_speedup}, the locality exploiting 
optimizations make the kernel $20\%$ faster than the original OP2 version.
Again the best performance for \texttt{SpaceDiscretization} was reached with hte AOS 
layout, which is not optimal for the whole application -- overall the best total performance
can be reached with SOA layout. Notably, \texttt{ComputeFluxes} also benefits significantly from the locality optimization, since this kernel also iterates on the edges and read data from cells; we experienced $5\%$ speed increase in \texttt{ComputeFluxes}.
The increased speed of the two kernels results in a $13\%$ increase for the whole
application. The useful bandwidth also increases and reached $30\%$ of the peak
stream bandwidth of the P100 GPU.

On the V100, the performance of the kernel was $462$ GB/s ($62\%$ of the peak bandwidth) with $18\%$ (AOS) speedup compared to the original implementation. Running the whole application with the reordered mesh, SOA 
achieved $11\%$ speedup on \texttt{SpaceDiscretization} while improving
\texttt{ComputeFluxes} with $10\%$ as well, leading to a $8\%$ speedup on the
whole application compared to the original implementation.

\begin{table}[Htbp]
\centering
\resizebox{\columnwidth}{!}{
\small
\begin{tabular}{|r|cc|c|c||c|}\hline
Reordering  & \multicolumn{2}{c|}{none} & \multicolumn{2}{c||}{partition} & 
original\\\hline
Data layout &    AOS      &   SOA     &     AOS      &    SOA      &    
SOA\\\hline
Bandwidth (GB/s) &   $ 133$ &   $ 120$ &   $ 146$ &   $ 134$ & $119$\\
Runtime (ms) & $0.87$ & $0.95$ & $0.77$ & $0.85$ & $0.93$\\
Achieved Occupancy &   $0.82$ &   $0.81$ &   $0.80$ &   $0.80$  &   $0.68$\\
Global Memory Read Transactions &   $ 9114\text{k}$ &   $ 9166\text{k}$ &   $ 
7493\text{k}$ &   $ 7617\text{k}$ &  $9504\text{k}$ \\
Global Memory Write Transactions &   $ 2438\text{k}$ &   $ 2512\text{k}$ &   $ 
1542\text{k}$ &   $ 1640\text{k}$ &   $ 2809\text{k}$ \\
Number of Block Colours &   $       5$ &   $       5$ &  $       9$ &   $       
9$ &   $       6$ \\
Number of Thread Colours &   $       3$ &   $       3$ &  $       4$ &   $       
4$ &   $       3$ \\
Reuse factor &   $     1.5$ &   $     1.5$ &  $     2.8$ &   $     2.8$ &   $    
 1.48$ \\
Issue Stall Reasons (Synchronization) &     $11\%$ &     $12\%$ &  $15\%$ &     
$15\%$ &     $14\%$ \\
Issue Stall Reasons (Data Request) &     $51\%$ &     $50\%$ &  $46\%$ &     
$46\%$ &     $47\%$ \\
Average cache lines/block &   $     300$ &   $     307$ &  $     165$ &   $     
184$ &   -  \\
Warp Execution Efficiency &   $  98\%$ &   $  98\%$ &   $  97\%$ &   $  97\%$ &  
 $  65\%$  \\\hline
Block size &  \multicolumn{4}{c||}{$307$} & $128$\\\hline
\end{tabular}
}
\caption{Low-level performance metrics of Volna's \texttt{SpaceDiscretization} 
kernel - hierarchical coloring. The last column details the measured performance 
of the original code.}
\label{tab:volna_counters_hier}
\end{table}

\subsubsection{BookLeaf}
\noindent The third application we explore is BookLeaf. It is a 2D unstructured
mesh Lagrangian hydrodynamics application from the UK Mini-App Consortium
\cite{uk-mac}. It uses a low order finite element method with an arbitrary
Lagrangian-Eulerian method.  BookLeaf is written entirely in Fortran 90 and has
been ported to use the OP2 API and library. BookLeaf has a large number of
kernels with different access patterns such as indirect increments similar to
increments inside \texttt{res\_calc} in Airfoil. For benchmarking we used the
SOD test case with a mesh of 4 million cells. The top time consuming kernel with
indirect increments is \texttt{getacc\_scatter}, which iterates on cells while
incrementing data indirectly on vertices, $6\%$ of total execution time is spent
in this kernel. This kernel reads 17 double values directly, and increments 4
double values on each of the 4 indirectly accessed nodes in each iteration.

\begin{figure}[Htbp]
\centering
\includegraphics[width=10cm]{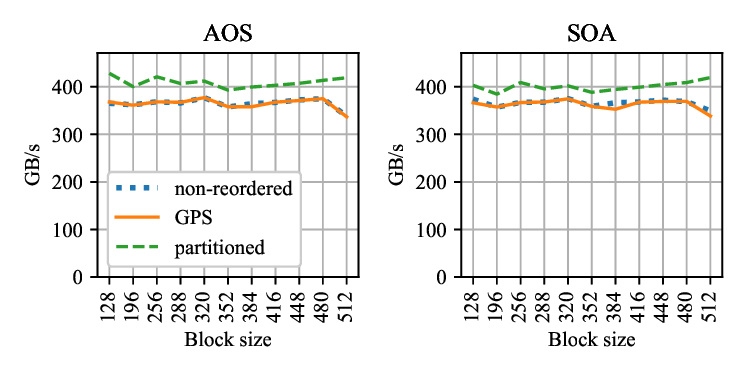}
\caption{Bookleaf's \texttt{getacc\_scatter} kernel bandwidth on a mesh with 
$4000000$ edges using hierarchical colouring.}
  \label{fig:bookleaf_bw-vs-bs_hier}
\end{figure}

\begin{figure}[Htbp]
\centering
\includegraphics[width=10cm]{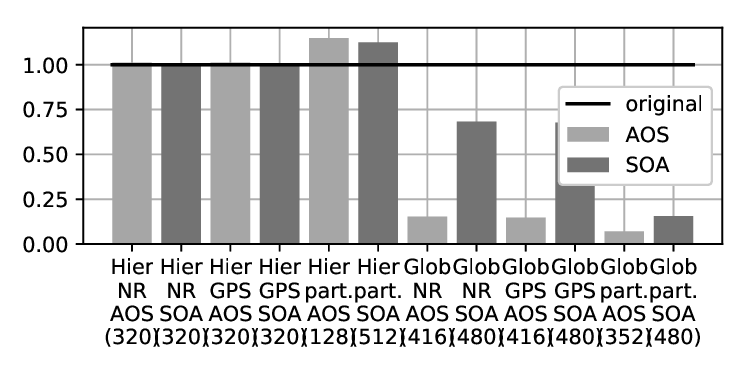}
\caption{Bookleaf's \texttt{getacc\_scatter} kernel speedup compared to the 
original code, done on a mesh with $4000000$ edges. The block sizes are shown 
in parentheses, the reordering algorithms are: the original reordering (NR), 
GPS reordering and partitioning (part.)}
  \label{fig:bookleaf_speedup}
\end{figure}

Runtime performance of BookLeaf and specifically the low-level performance of 
the \texttt{getacc\_scatter} kernel are detailed in Figures
\ref{fig:bookleaf_bw-vs-bs_hier} and \ref{fig:bookleaf_speedup}. Again we see  
benefits from partitioning. The register count and occupancy are also similar to 
those with Airfoil ($64$ registers, achieving occupancy around $40\%$), this 
now leads to the variations in performance for different block sizes. With 
partitioning, the number of thread colors increases from $2$ to $5$, this
leads to increased stalls from synchronizations: from $9\%$ to $20\%$, while
the reuse factor increases (from $2$ to $3.5$). This is comparable to 
that of Airfoil, and explains the smaller increase in performance (only $15\%$, 
compared to the $19\%$ increase in Airfoil). The higher data reuse leads to 
$14\%$ and $41\%$ decrease of the number of global transactions, for reads and 
writes, respectively. Such a large difference between reads and writes is also 
due to \texttt{getacc\_scatter} having no indirect reads, similar to 
like \texttt{SpaceDiscretization} in Volna. 

The best performance we achieved with \texttt{getacc\_scatter} ($1.15\times$ 
speedup) results in about $1\%$ performance increase for the the whole 
application. The useful bandwidth also increased and reached $83\%$ of the peak 
stream bandwidth of the P100 GPU.
On the V100, the performance of the kernel was $631$ GB/s ($85\%$ of the peak
bandwidth) which results in $0.5\%$ runtime speedup on the whole application
compared to the original.

\subsubsection{LULESH}\label{sec:lulesh-summary}

\hyphenation{CalcFBHourglassForceForElems IntegrateStressForElems}

\noindent LULESH (Livermore Unstructured Lagrangian Explicit Shock 
Hydrodynamics~\cite{LULESH2:changes}) application represents a typical 
hydrocode representing the Shock Hydrodynamics Challenge Problem that was 
originally defined and implemented by Lawrence Livermore National Lab as one of 
five challenge problems in the DARPA UHPC program. It  has since become a widely 
studied proxy application in DOE co-design efforts for exascale. 

LULESH is a highly simplified application, hard-coded to only solve a simple
Sedov blast problem that has an analytic solution~\cite{LULESH:spec} – but
represents the numerical algorithms, data motion, and programming style typical
in scientific C/C++ based applications at the Lawrence Livermore National
Laboratory. LULESH approximates the hydrodynamics equations discretely by
partitioning the spatial problem domain into a collection of volumetric elements
defined by a mesh. The mesh itself is structured (and generated in the code), 
but the algorithm doesn't take this into account and accesses the data through 
an eight-dimensional mapping for the hex8 (brick) elements.

We explore the \texttt{IntegrateStressForElems} kernel that calculates the
forces in the nodes. Iterating over cells, it reads 3 double values from each 
of the 8 indirectly accessed nodes, increments 3 double values for each node, 
reads 3 double values directly and writes 1 double value directly. In our 
measurements, we used a mesh with \num{4913000} cells and \num{5000211}
nodes. The original CUDA version of the code contracted this kernel with
\texttt{CalcFBHourglassForceForElems}; the only modifications we applied to 
this code for our tests was to remove these parts from the kernel.

The \texttt{IntegrateStressForElems} kernel uses a mapping with 8 neighbors for 
the brick elements (compared to the 2--4 as in the case of the previous 
application kernels). As a result, the number of block colors is quite high: $8$, $16$ 
and $24$ in the global coloring versions (for the different reorderings), and 
$4$, $5$ and $15$ in the hierarchical coloring versions. The number of thread 
colors was also quite high: $4$ in the non-reordered ($4.5$ in the GPS) and 
$11.6$ in the partitioned version. The non-reordered and GPS versions yield 
blocks that have ``pencil shape'', thus requiring fewer thread colors, whereas 
the partitioned version yields more cubical shaped blocks, leading to the higher 
number of thread colors. This is a much larger increase compared to the previous 
applications (Table~\ref{tab:lulesh_counters_hier}). At the same time of course, 
data reuse is higher compared to 2D applications - between $2.6$ and $4.8$.

The other aspect in which LULESH is differs is that it uses a high amount of
registers ($96$), which significantly decreases occupancy: with block size 320, 
the AoS version achieved $15\%$ and the SoA version achieved around $30\%$. 
Because of these two reasons, the synchronization overhead ($39\%$ stalls were 
from synchronization on the partitioned mesh) couldn't be hidden: there were no 
warps from other blocks to be scheduled in place of the stalled ones because
there was only one block running on each multiprocessor. The difference in 
achieved occupancy also means that the SoA version with two blocks 
per multiprocessor gives better performance.

\begin{figure}[Htbp]
\centering
\includegraphics[width=10cm]{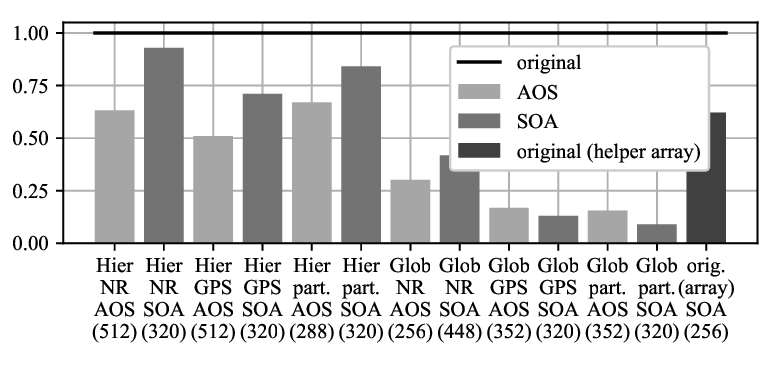}
\caption{LULESH's \texttt{IntegrateStressForElems} kernel speedup compared to
  the original (gathering) code, done on a mesh with $4913000$ cells. The block
  sizes are shown in parentheses, the reordering algorithms are: the original
  reordering (NR), GPS reordering and partitioning (part.) The last bar shows
  the relative performance of the original code with the helper array
  approach.}
\label{fig:lulesh_speedup}
\end{figure}

The original implementation used either atomics or helper arrays as a means of
avoiding data races. As shown in Figure \ref{fig:lulesh_speedup}, the
hierarchical coloring algorithm performs significantly better (giving a $49\%$
speedup) than the original two-step implementation (and also uses much less
memory), but it is worse than the original atomics implementation by $17\%$.

\begin{figure}[Htbp]
\centering
\includegraphics[width=10cm]{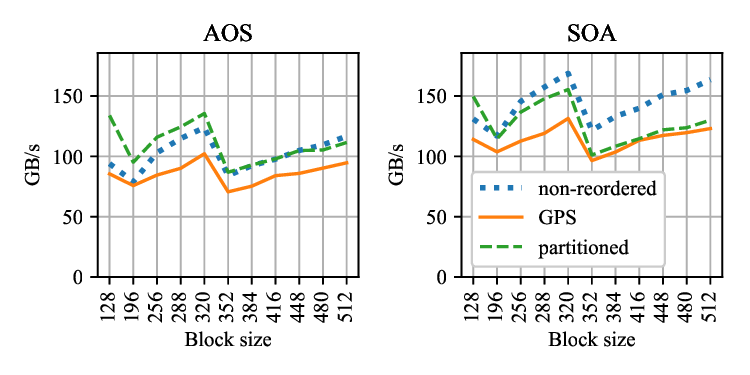}
\caption{LULESH's \texttt{IntegrateStressForElems} kernel bandwidth on a mesh 
with $4913000$ cells using hierarchical colouring.}
  \label{fig:lulesh_bw-vs-bs_hier}
\end{figure}

In the original application version $37\%$ out of the total time is spent in
the \texttt{IntagrateStressForElems} kernel, therefore the achieved $49\%$
speedup on the kernel over the two-step implementation gives about $11\%$ (for
our best result) or $8\%$ (for our partitioned version) speedup on the whole
application. This was measured by reading back the reordered mesh in the
original code for all kernels. The useful bandwidth in case of the best version
of \texttt{IntegrateStressForElems} reached $35\%$ of the peak stream bandwidth
on the P100 GPU.

On the V100, we achieved $51\%$ speedup for this kernel compared to the original
two-step implementation with $270$ GB/s bandwidth ($36\%$ of the peak
bandwidth). In the original version $38\%$ of the total time is spent in
\texttt{IntegrateStressForElems}, therefore this achieved kernel speedup results
in about $16\%$ speedup on the whole application. Again, the atomics version
performed best with a bandwidth of $561$ GB/s ($76\%$ of peak bandwidth), two
times faster than our version.

\begin{table}[Htbp]
  \centering
  \resizebox{\columnwidth}{!}{
    \small
  \begin{tabular}{|r|cc|c|c||c|}
    \hline
      Reordering  & \multicolumn{2}{c|}{none} & \multicolumn{2}{c||}{partition} 
& original\\
      \hline
      Data layout &    AOS      &   SOA     &     AOS      &    SOA & SOA\\
      \hline
                               Bandwidth (GB/s) &   $ 80$ &   $ 172$ &   $ 83$ 
&   $ 157$ & $97$\\
                               Runtime (ms)     &   $8.48$ & $3.90$ & $8.10$ & 
$4.31$ & $6.98$ \\
                             Achieved Occupancy &   $0.15$ &   $0.29$ &   $0.15$ 
&   $0.30$ & $0.24$\\
        Global Memory Read Transactions (total) &   \num{33546}k & \num{35395}k 
&   \num{22883}k &   \num{25006}k  & \num{16072}k\\
       Global Memory Write Transactions (total) &   \num{12674}k & \num{12706}k 
&   \num{ 8052}k &   \num{ 8703}k  & \num{32689}k\\
                        Number of Block Colours &   $       4$ &   $       4$ & 
$      15$ &   $      15$ & - \\
                       Number of Thread Colours &   $       4$ &   $       4$ & 
$    9.8$ &   $    9.8$  & -\\
                                   Reuse Factor &   $     2.6$ &   $     2.6$ & 
$     4.8$ &   $     4.8$ & -\\
          Issue Stall Reasons (Synchronization) &   $  13\%$ &   $  19\%$ &   $ 
36\%$ &   $  39\%$  & $0\%$\\
             Issue Stall Reasons (Data Request) &   $  63\%$ &   $  56\%$ &   $ 
38\%$ &   $  34\%$  & $26\%$\\
                      Average Cache Lines/Block &   $     744$ &   $     747$ & 
$     427$ &   $     474$  & -\\
                      Warp Execution Efficiency &   $  98\%$ &   $  98\%$ &   $ 
94\%$ &   $  93\%$ & $100\%$\\
      \hline
      Block size & \multicolumn{4}{c||}{$320$} & $64$\\

    \hline
  \end{tabular}
  }
\caption{Low-level performance metrics of LULESH's 
\texttt{IntegrateStressForElems} kernel - hierarchical coloring. The last column 
details the measured performance of the original kernel.}  
\label{tab:lulesh_counters_hier}
\end{table}

\subsubsection{miniAero}\label{sec:mini-aero-summary}

The final application we explore is miniAero \cite{miniaero}, which is a 
mini-application from the Mantevo suite \cite{heroux2009improving}. MiniAero is 
an explicit (4th order Runge-Kutta) unstructured finite volume code that solves 
the compressible Navier-Stokes equations. Both inviscid and viscous terms are 
included. The viscous terms can be optionally included or excluded. For 
miniAero, meshes are created within the code and are simple 3D hex8 meshes. 
These meshes are generated on the CPU and then moved to the GPU. While 
the meshes generated in code are structured, the code itself uses unstructured 
mesh data structures and access patterns. This mini-application uses the Kokkos 
library~\cite{CarterEdwards20143202}.

For miniAero we study the \texttt{compute\_face\_flux} kernel that computes the
flux contributions of the faces and increments it with the appropriate cell flux
values. The kernel iterates over the faces of the mesh, and accesses the cells
indirectly. In each iteration, it reads 28 and increments 5 double values from
each of the 2 indirectly accessed cells, and reads 12 double values directly. 
The original code, depending on a compile time parameter, either uses 
the auxiliary \texttt{apply\_cell\_flux} kernel that does the actual 
incrementing by gathering the intermediate results from a large temporary array, 
or uses atomics to do it within the kernel. Both the atomics and the work of 
the auxiliary kernel was substituted in our code by coloring.

The \texttt{compute\_face\_flux} kernel is the most computationally intensive
among the ones we tested: it uses $165$ registers in hierarchical coloring
($166$ in SoA layout). Also, it achieves, with a block size of $384$ and 
reordered by GPS, $15\%$ of peak double precision efficiency, compared to the 
$6$--$7\%$ in Airfoil (Table \ref{tab:mini_aero_counters_hier}). It also uses 
$8$ square root operations and several divides that can't efficiently fill the 
pipelines at such low occupancy. 

\begin{figure}[Htbp]
\centering
\includegraphics[width=10cm]{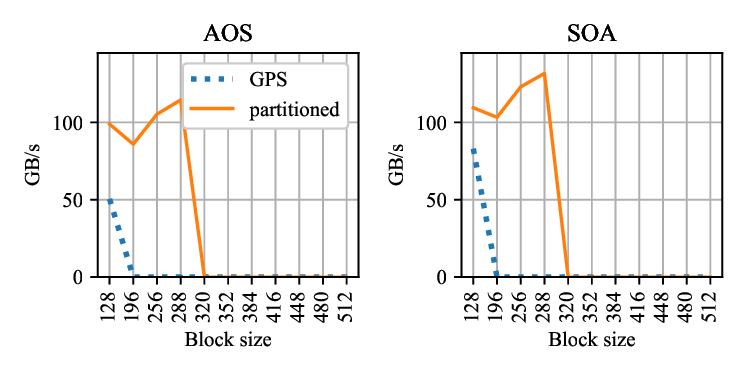}
\caption{miniAero's \texttt{compute\_face\_flux} kernel bandwidth on a mesh 
with $6242304$ faces. The kernel didn't run in the cases where the data reuse 
was not high enough because the large amount of shared memory needed; these 
are shown here with $0$ bandwidth.}
  \label{fig:mini_aero_bw_crash}
\end{figure}

The amount of data indirectly accessed by the kernel is also significant: each 
thread accesses $2$ data points indirectly, each holding $32$ double precision 
values. If all of these values are loaded into shared memory, the size exceeds 
the hardware limits with block sizes larger than $288$; it didn't run with the 
original mesh numbering with any block size, and only with smaller block sizes 
on the reordered meshes (Figure \ref{fig:mini_aero_bw_crash}). The other 
measurements were carried out by only loading the incremented data into shared 
memory.

\begin{figure}[Htbp]
\centering
\includegraphics[width=10cm]{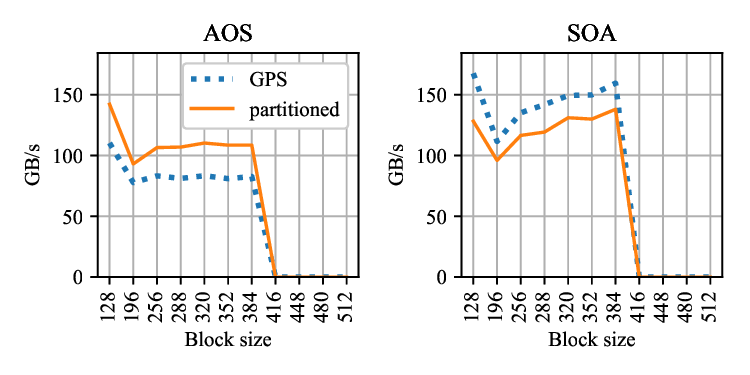}
\caption{miniAero's \texttt{compute\_face\_flux} kernel bandwidth on a mesh 
with $6242304$ faces. The shared memory was only used to cache the increments,
reducing the need for large shared memory size. The kernel didn't fit into the
shared memory with block sizes larger than $384$ or if not reordered because
the large amount of shared memory needed; these are shown here with $0$
bandwidth.} \label{fig:mini_aero_bw_small-cache}
\end{figure}

The mesh also has a complex structure. $18$ and $15$ block colors for the GPS
reordered and partitioned versions, respectively. The original ordering was
far from optimal, we couldn't run the non-reordered version, because the 
number of block colors exceeded the implementation limit of the library, which 
is $256$. As with LULESH, only one block was running at a time on each 
multiprocessor. Although the synchronization overhead was lower ($3$ and $6$ 
thread colors in the GPS reordered and partitioned versions, respectively), the 
costly operations prevented high performance gains in the case of the 
partitioned version (Figure \ref{fig:mini_aero_bw_small-cache}).

\begin{figure}[Htbp]
\centering
\includegraphics[width=10cm]{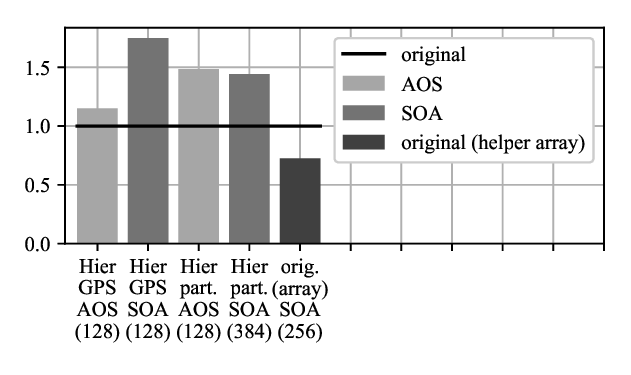}
\caption{miniAero's \texttt{compute\_face\_flux} kernel speedup compared to 
the original code that was using atomic adds, on a mesh with $6242304$ faces. 
The shared memory was only used to cache the increments, reducing the need 
for large shared memory size. The last bar shows the relative performance of 
the original code with the helper array approach.}
\label{fig:mini_aero_speedup_small-cache}
\end{figure}

The original Kokkos implementation either used atomic adds or the two-step
gathering approach depending on compilation parameters. Our implementation
outperformed both with a 75\% speedup (Figure
\ref{fig:mini_aero_speedup_small-cache}). The useful bandwidth in case of the best version of
\texttt{compute\_face\_flux} ($168$ GB/s) reached $34\%$ of the peak stream
bandwidth of the P100 GPU.

\begin{table}[Htbp]
  \centering
  \resizebox{\columnwidth}{!}{
    \small
  \begin{tabular}{|r|cc|c|c||c|}
    \hline
      Reordering  & \multicolumn{2}{c|}{GPS} & \multicolumn{2}{c||}{partition} & 
original\\
      \hline
      Data layout &    AOS      &   SOA     &     AOS      &    SOA & SOA\\
      \hline
                               Bandwidth (GB/s)&    $  67$ &   $ 84$ &   $ 117$ 
&   $ 113$ & $96$\\
                               Runtime (ms)    &    $19.09$ & $15.38$ & $11.04$
                               & $11.38$ & $13.39$\\
                             Achieved Occupancy&    $0.06$ & $0.06$ & $0.12$ & 
$0.12$  & $0.23$\\
                Global Memory Read Transactions&    $73561\text{k}$ & 
$82028\text{k}$ & $53403\text{k}$ & $77091\text{k}$  & $87368\text{k}$\\
               Global Memory Write Transactions&    $9153\text{k}$ & 
$10390\text{k}$ & $6694\text{k}$ & $9008\text{k}$ & $5934\text{k}$ \\
                        Number of Block Colours&    $      18$ &   $      18$ & 
$      15$ &   $      15$  & -\\
                       Number of Thread Colours&    $       3$ &   $       3$ & 
$       6$ &   $       6$  & -\\
                                   Reuse Factor&    $     2.2$ &   $     2.2$ & 
$     3.9$ &   $     3.9$  & -\\
          Issue Stall Reasons (Synchronization)&    $4\%$ & $9\%$ & $15\%$ & 
$21\%$    & $0\%$ \\
             Issue Stall Reasons (Data Request)&    $61\%$ & $35\%$ & $47\%$ & 
$35\%$  & $49\%$\\
     Issue Stall Reasons (Execution Dependency)&    $23\%$ & $33\%$ & $23\%$ & 
$23\%$  & $13\%$\\
                      Average Cache Lines/Block&    $     452$ &   $     471$ & 
$     269$ &   $     344$  & -\\
                      Warp Execution Efficiency&    $91\%$ & $84\%$ & $88\%$ & 
$85\%$  & $91\%$\\
                   FLOP Efficiency(Peak Double)&    $5\%$ & $8\%$ & $10\%$ & 
$12\%$  & $8\%$\\
      \hline
      Block size & \multicolumn{4}{c||}{128} & 256\\
    \hline
  \end{tabular}
  }
\caption{ Low-level performance metrics of the miniAero's 
\texttt{compute\_face\_flux} kernel - hierarchical coloring. The last column 
details the measured performance of the original code.}
  \label{tab:mini_aero_counters_hier}
\end{table}

On Volta, we achieved $158\%$ (compared to the atomic version) and $92\%$
(compared to the two-step version) speedups in the kernel $268$ GB/s bandwidth
($36\%$ of the peak bandwidth).

\subsubsection{Analysis of structured meshes}
\label{analysis-of-structured-meshes}

\noindent Recalling that the meshes of miniAero and LULESH are actually 
structured meshes, generated by the code itself allows us to use their 
structured nature to create partitions with better shapes as opposed to 
partitions produced by METIS. This in turn allows us to understand the trade-off 
between high data reuse and number of thread colors.  We create 1D (straight 
line), 2D (rectangles) and 3D (bricks) shape partitions such that these will 
have an increasing amount of reuse, and with that, number  of colors. 

\begin{figure}[Htbp]
\centering
\includegraphics[width=12cm]{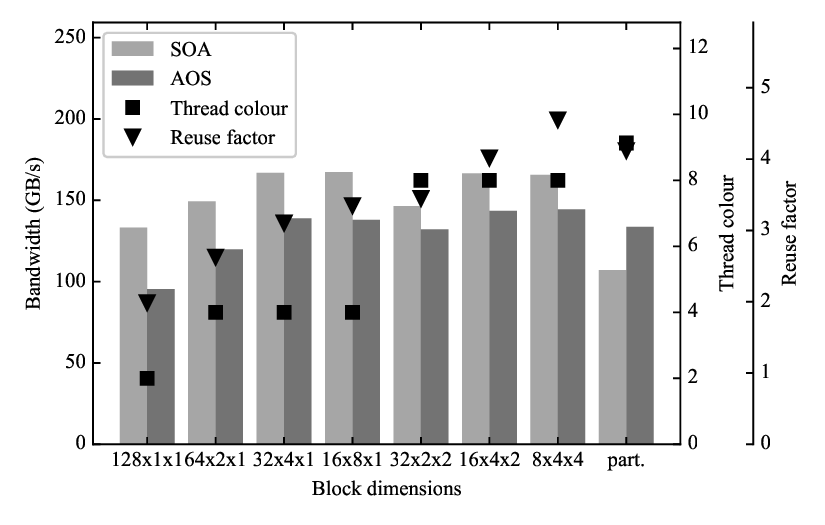}
\caption{LULESH's \texttt{IntegrateStressForElems} kernel with explicitly 
controlled partitioning. For comparison, the last column shows the result on 
the same mesh, partitioned by METIS.} \label{fig:lulesh_block}
\end{figure}

\begin{figure}[Htbp]
\centering
\includegraphics[width=12cm]{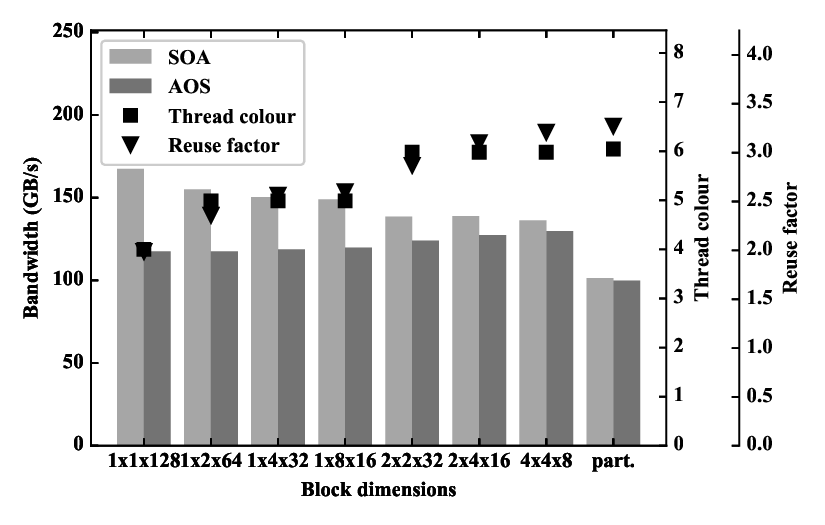}
\caption{miniAero's \texttt{compute\_face\_flux} kernel with explicitly 
controlled partitioning. For comparison, the last column shows the result on 
the same mesh, partitioned by METIS.} \label{fig:mini_aero_block}
\end{figure}

While both kernels operate on 3D Cartesian (hex8) meshes, the LULESH kernel 
\texttt{IntegrateStressForElems} uses a mapping from cells to their connected 
vertices, and the \texttt{compute\_face\_flux} kernel in miniAero maps from 
(internal) faces to cells. We created a number of different partition shapes - 
1D lines, 2D rectangles and 3D bricks. Figures~\ref{fig:lulesh_block} and 
\ref{fig:mini_aero_block} show the bandwidths, reuse factors and the number of 
thread colors across different block-shapes, along with the result of 
partitioning the same mesh using METIS. The size of the blocks is $128$. The 
results are from meshes with specifically tailored shapes so that the 
handcrafted blocks can cover them without any gaps. 

In \texttt{IntegrateStressForElems}, the achieved bandwidth of the 
original ordering is $134\,\text{GB/s}$ - it uses a row major order. This is 
similar to what we use for a line block shape that is $128$ cells long and only $1$ 
cell thin in the other dimensions. The partitioned mesh (as detailed in Section 
\ref{sec:lulesh-summary}) achieved $132\,\text{GB/s}$, and we achieved 
$167\,\text{GB/s}$ bandwidth using our handcrafted blocks, for the same block 
size ($128$). Note that using regular shapes is better for the thread coloring 
algorithm as well. With METIS partitioning, the number of colors needed 
is higher than in the other cases.

For \texttt{compute\_face\_flux}, we achieved $167\,\text{GB/s}$ bandwidth,
compared to $101\,\text{GB/s}$ achieved with METIS partitioning. Of course, 
using these handcrafted blocks can only be done on meshes that are actually 
structured, therefore this is not representative of realistic cases. However, 
these results illustrate clearly that when the number of thread colors are the 
same, increased reuse leads to better performance. Also, there is an optimal 
trade off between reuse and the number of thread colors for each application, 
and performance will suffer above/below that. As an alternative to METIS, we 
explored using SCOTCH~\cite{Pellegrini1996} to create the required partitions. 
However SCOTCH also did not produce a better partitioning for these 
applications. The challenge lies in finding a partitioning algorithm that can 
either find the middle ground, or can be tuned  along the amount of reuse it 
aims to achieve. Such partitioning algorithms are not currently available, but 
the evidence here clearly demonstrate their usefulness.

\section{Conclusion}\label{conclusion}

\noindent In this work we presented a number of novel locality-exploiting 
optimizations for the efficient execution of unstructured-mesh algorithms on 
GPUs. The key focus was to improve performance of kernels with indirect 
increment data-access patterns. We build on well known techniques such as 
data-layouts (AoS and SoA), graph bandwidth minimizing algorithms, global and 
hierarchical coloring approaches for exploring efficient execution strategies. 
A novel reordering algorithm which uses k-way recursive partitioning, together 
with the use of GPU shared memory implementing a hierarchical coloring method is 
designed to significantly improve data reuse within CUDA thread blocks.

The new optimizations are then applied to several well established unstructured
mesh applications, investigating their performance on NVIDIA’s latest P100
and V100 GPUs. A range of performance metrics were benchmarked including 
runtime, memory transactions, achieved bandwidth performance, GPU occupancy and 
data reuse factors and are used to understand and explain the key factors 
impacting performance.

When comparing the performance of global coloring to that of hierarchical 
coloring (with shared memory), we demonstrated that the latter approach 
consistently performed better. This was due to its ability to exploit the 
temporal locality in indirectly accessed data by avoiding data races in shared 
memory with synchronization within thread blocks rather than different kernel 
launches.

Analyzing the performance of reordering based on GPS renumbering and 
partitioning showed that former improves global coloring with increasing 
spatial reuse, while the latter can significantly improve the shared memory 
approach by increasing data reuse within thread blocks. In this case, we 
see smaller shared memory usage and fewer global memory transactions.

We also see that there is a trade-off between high data reuse and large numbers 
of thread colors in hierarchical coloring. This is especially pronounced in 3D 
applications, and when the achieved occupancy is low: the more thread colors a 
block has, the more synchronizations it will need, the latency of which can be 
hard to hide when there are few eligible warps.

The locality exploiting optimizations detailed in this paper enable us to 
improve performance of indirect kernels by $19\%$ on Airfoil, $20\%$ on 
Volna, $15\%$ on Bookleaf, $75\%$ on Lulesh and $75\%$ on miniAero over the 
original implementations, on the GPUs tested. These results advances the state 
of the art, demonstrating that the algorithmic patterns used in most current 
implementations (particularly in case of US DoE codes represented by 
LULESH and MiniAero) could be significantly improved upon by the adoption of 
two-level coloring schemes and partitioning for increased data reuse.

When carrying out this work, it had become clear that partitioning algorithms in
traditional libraries such as METIS and SCOTCH were not particularly well 
suited for producing such small partition sizes. As potential future work, we 
wish to explore algorithms that are better optimized for this purpose. The 
performance of these partitioning algorithms was also low - parallelizing this 
could be another interesting challenge. Finally, we are planning to integrate 
these algorithms into the OP2 library, so they can be automatically deployed on 
applications that already use the OP2 library, such as Airfoil, BookLeaf, Volna 
or Rolls-Royce Hydra.

The optimized algorithms are implemented as an open-source software 
library~\cite{opt-library}. 

\section*{Acknowledgements}
\noindent Istv\'an Reguly was supported by the J\'anos Bolyai Research 
Scholarship of the Hungarian Academy of Sciences. Project no. PD 124905 has been 
implemented with the support provided from the National Research, Development 
and Innovation Fund of Hungary, financed under the PD\_17 funding scheme. Gihan 
Mudalige was supported by the Royal Society Industrial Fellowship Scheme 
(INF/R1/180012). The authors would like to acknowledge the use of the University 
of Oxford Advanced Research Computing (ARC) facility in carrying out this work 
\url{http://dx.doi.org/10.5281/zenodo.22558}. The research has been supported 
by the European Union, co-financed by the European Social Fund 
(EFOP-3.6.2-16-2017-00013).
%
%
\bibliographystyle{elsarticle-num}
\bibliography{bibliography}

\end{document}